\documentclass[a4paper]{PoS}
\pdfoutput=1
\usepackage{amsmath,amssymb,amsbsy,multicol,url}
\usepackage[utf8]{inputenc}
\usepackage[numbers,sort&compress]{natbib}
\usepackage{natbibspacing}
\usepackage{graphicx}
\usepackage{ulem,color}
\newcommand{\red}[1]{\textcolor{red}{ #1}}

\title{Infrared properties of a prototype  model for beyond-Standard Model physics\footnote{This article combines the contribution ``Infrared properties of a prototype pNGB model for BSM physics"  by Anna Hasenfratz and the contribution ``Spectrum of a prototype model with the Higgs as pNGB'' by Claudio Rebbi.}}

\ShortTitle{Infrared properties of a model for beyond-Standard Model physics}

\author{\speaker{Anna Hasenfratz} \\
        Department of Physics, University of Colorado, Boulder, CO, USA\\
        E-mail: \email{Anna.Hasenfratz@colorado.edu}}

\author{Claudio Rebbi$^{\ddagger}$\\ 
        Department of Physics, Boston University, Boston, MA, USA \\
        E-mail: \email{rebbi@bu.edu}}

\author{Oliver Witzel \\
  Higgs Centre for Theoretical Physics, School of Physics \& Astronomy,\\
  The University of Edinburgh, Edinburgh, United Kingdom \\
        E-mail: \email{o.witzel@ed.ac.uk}}

\abstract{
We construct a prototype BSM model based on the SU(3) color gauge group and a combination of 4 light (massless) and 8 heavy flavors. In the infrared, the SU(4) flavor chiral symmetry is spontaneously broken, while in the ultraviolet this model exhibits the properties of the $N_f=12$ conformal fixed point.
Renormalization group considerations predict the spectrum of such a system to show hyperscaling, i.e.~dimensionless ratios of hadron masses or decay constants are independent of the heavy mass. Hyperscaling is present for bound states of light, heavy, or a combination of heavy and light flavors and leads to a strongly predictive model. Despite chiral symmetry breaking, this system features a spectrum exhibiting a very non-QCD like behavior.  Furthermore, the gauge coupling becomes an irrelevant parameter.
We support these expectations by presenting numerical results based on four different values of the heavy quark mass $am_h$, up to six different values of the light quark mass $am_\ell$, and include, for the first time, preliminary data at a second value of the gauge coupling $\beta$. Our model can be embedded in scenarios describing the Higgs boson either as a pseudo Nambu-Goldstone boson or a dilaton-like particle.  
}

\FullConference{34th annual International Symposium on Lattice Field Theory\\
		 24-30 July 2016\\
		 University of Southampton, UK}

\begin{document}

\section{Introduction}
\label{sec:introduction}

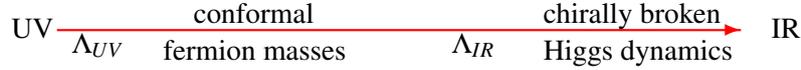
\begin{figure}[!h]
  \centering
\begin{picture}(105,10)
\put(0,5){UV}    \put(6,6){\thicklines\red{\vector(1,0){90}}} \put(100,5){IR}
\put(8,3){$\Lambda_{UV}$} \put(58,3){$\Lambda_{IR}$}
\put(24,7){conformal} \put(20,2){fermion masses}
\put(70,7){chirally broken} \put(70,2){Higgs dynamics}
\end{picture}
\label{fig1}
\caption{Schematic representation of the transition between UV and IR 
regimes in strong dynamics models for EW symmetry breaking.}
\end{figure}

While the analysis of data accumulated by ATLAS and CMS in 2016 do
not provide compelling evidence for beyond-Standard Model (BSM) phenomena \cite{ICHEP2016_LHC_highlights1, ICHEP2016_LHC_highlights2},
it is generally accepted that the Standard Model (SM) is only an effective theory.
New interactions are needed for an ultraviolet (UV) completion of the
Higgs sector, which otherwise leads to a triviality in the UV, as well
as to explain other phenomena, such as dark matter, for which there
is tantalizing evidence  but no established theory.  
Many of the models which are currently
considered for BSM physics are based on a large scale separation between 
the infrared (IR) and ultraviolet (UV) regimes~\cite{Contino:2010rs,Luty:2004ye,Dietrich:2006cm,Brower:2015owo,Csaki:2015hcd,Arkani-Hamed:2016kpz},
that leads to a ``walking'' gauge coupling and provides a dynamical mechanism
for electroweak (EW) symmetry breaking. This mechanism avoids tuning of the
Higgs mass within a framework which  may satisfy EW precision 
measurement constraints.  Figure~\ref{fig1} illustrates the assumed transition between the UV and IR regimes.
An almost universal features of all strongly coupled BSM models is that they assume
the proximity of an IR fixed point, which is eventually responsible for the
walking of the gauge coupling.  This in turn requires the presence
of a sufficient number of fermions to push the theory close to
the conformal window, whose presence is indicated by perturbative \cite{Ryttov:2010iz,Ryttov:2016hal}
as well as numerical arguments (for recent references see \cite{DeGrand:2015zxa,Nogradi:2016qek}).  In order to investigate the effect
which a large number fermions will have on the transition between the
UV and IR regimes, in collaboration with Richard Brower and
Evan Weinberg, we embarked in the numerical study of an SU(3)
theory with four light fermions and eight heavy fermions of variable
mass~\cite{Hasenfratz:2016gut,Brower:2015owo,Hasenfratz:2015xca,Weinberg:2014ega,Brower:2014ita,Brower:2014dfa}  The original goal was 
to see how, through the variable mass, one could interpolate from
the confining behavior of a QCD-like theory to the conformal behavior
of a theory with 12 flavors of (almost) massless fermions.  The investigation
led to some  interesting, unique results:  While we did observe a changing
behavior of the renormalized coupling, which exhibited a growing
region of slow walking as the mass of the eight heavy fermions decreased,
the spectrum of light composites with a properly rescaled lattice
spacing did not appear to change with the heavy fermions mass. A
result which we interpreted as evidence of hyperscaling, a consequence of infrared conformal behavior  and in agreement with step scaling studies of the 12-flavor system~\cite{Hasenfratz:2016dou}.  In these 
proceedings we report on an extension of that early work, 
where we examine in detail the spectrum of heavy composites, 
finding results which further confirm hyperscaling and may have implications 
for future observations at the LHC.  We also performed simulations 
at a different value of the gauge coupling constant to validate the 
notion that the gauge coupling is an irrelevant parameter in the 
neighborhood of an IR fixed point.

\section{Hyperscaling}
\label{sec:hyperscaling}

We start from the premise that in the limit where all twelve flavors
are massless our theory has an infrared fixed point (IRFP).   This assumption has substantial numerical support \cite{Cheng:2013eu,Itou:2013ofa,Itou:2014ota,Cheng:2013xha,Lombardo:2014pda} and remaining concerns \cite{Fodor:2011tu,Fodor:2016zil} were recently addressed in 
Ref.~\cite{Hasenfratz:2016dou}.
In the basin of attraction of the IRFP a Wilsonian renormalization group
analysis can be used to investigate  how  changes in the bare
fermion masses effect physical observables.  Working with the lattice regularization and
inside the conformal window, the bare parameters of the theory can be separated
into irrelevant gauge couplings $g_i$ and relevant lattice masses 
$\widehat m_i = a m_i$. The critical surface is given by $\widehat m_i = 0$ 
where the system is conformal at the IRFP  $g_i^\star$.

In the neighborhood of the IRFP an RG transformation that changes the
scale $\mu \to \mu^\prime = \mu /b $ ($b>1$) drives the gauge
couplings to $g_i^\star$, while masses transform with the scaling
dimension $y_m=1+\gamma_m$ as $\widehat m_i \to \widehat m_i^\prime =
b^{y_{m}} \widehat m_i$.
The correlation function of an operator $H$, after rescaling all
dimensional quantities by $b$, change as
 \begin{align}
C_H(t; g_i, \widehat m_i,\mu) =  b^{-2y _H} C_H(t/b; g_i^\prime, \widehat m_i^\prime,\mu)\, ,
\label{eq:C_H1}
 \end{align} 
where $y_H$ is the scaling dimension of $H$~\cite{DeGrand:2009mt,DelDebbio:2010ze}.

As $b$ increases the fermion mass increases and eventually the fermions 
decouple from the IR dynamics when the mass reaches the cutoff at
$\widehat m_i^\prime \sim \mathcal{O}(1)$.

This analysis can be extended to the case when there are two different
fermions masses $\widehat m_h = a m_h$ and $\widehat m_\ell = a
m_\ell$, as in our model, with $\widehat m_h \ge \widehat m_\ell$.
Since both masses scale with the same exponent $y_m$, the dependence
on $\widehat m_i^\prime = (\widehat m_h^\prime$, $\widehat
m_\ell^\prime)$ in Eq.~(\ref{eq:C_H1}) can be replaced with $(\widehat
m_h^\prime, \widehat m_\ell/ \widehat m_h) =( \widehat m_h^\prime,
m_\ell/ m_h ) $
  \begin{align}
C_H(t; g_i, \widehat m_i,\mu) =  b^{-2y _H} C_H(t / b ; g_i^\prime, \widehat m_h^\prime  , m_\ell /  m_h, \mu).
\label{eq:C_H1b}
 \end{align} 
The heavy fermions decouple when $\widehat m_h^\prime = b^{y_m}
\widehat m_h = \mathcal{O}(1)$.  Below that scale the dependence on
$\widehat m_h$ is through the ratio $m_\ell/m_h$. The infrared limit
for the light flavors can be set at $b = \widehat
m_\ell^{-1/y_m}$. Then Eq.~(\ref{eq:C_H1b}) reduces to
   \begin{align}
C_H(t; g_i, \widehat m_i,\mu) =  \widehat m_\ell^{2y _H/y_m} C_H(t \widehat m_\ell^{1/y_m} ; g_i^\prime ,  m_\ell /  m_h, \mu).
\label{eq:C_H1c} 
 \end{align} 
Correlation function behave exponentially at large distances,
  \begin{align}
 C_H(t; g_i,\widehat m_i,\mu) \propto e^{-M_H t}, \quad \quad t \to \infty.
 \label{eq:C_asymp} 
 \end{align}
 Comparison of the $t$ dependence in  Eqs.~(\ref{eq:C_H1c}) and (\ref{eq:C_asymp}) 
produces the scaling relation
 \begin{align}
 a M_H = (\widehat m_\ell)^{1/y_m}  F_H( m_\ell/ m_h).
\label{eq:M_scaling_b}
 \end{align}
Assuming that $b$ is large enough that the gauge couplings take their IRFP value, 
$g^\prime_i = g^\star_i$, $ F_H$ will be a  function of $ m_\ell/ m_h$ only, independent of the bare gauge couplings. 
Ratios of masses
  \begin{align}
\frac{M_{H1} }{M_{H2}} = \frac{ F_{H1}( m_\ell/  m_h)}{  F_{H2}( m_\ell/ m_h)} .
\label{eq:R_scaling}
 \end{align}
depend only on $ m_\ell/  m_h$, while the scaling function $F_{H1}/F_{H2}$ 
can of course be different for different observables.

As we will see in the next section, our results for light fermion composites 
and of heavy fermion composites, corroborate this prediction: 
composite masses and decay constants obtained with different $m_\ell$ and $m_h$
fall on universal curves as function of $ m_\ell/ m_h$. 
We expect the same to hold for the spectrum of heavy-light composites, 
but did not verify it by numerical simulations. 
Small deviations from universality can arise from corrections to
scaling due to the slowly running gauge coupling, i.e.~deviations from
$g_i^\prime = g_i^\star$. We have investigated these corrections
within the $N_f=12$ system and a similar analysis could be repeated
here~\cite{Cheng:2013xha}.  We have not done this, though, in our 
present work.

\section{Numerical results}
\label{sec:numerics}

We simulated an SU(3) system with one staggered field (= 4 flavors) with light mass 
$m_\ell$ and two  staggered fields (= 8 flavors) with heavy mass $m_h$.   
Simulations were done with $am_\ell=0.003,\, 0.005,$  $0.010,\, 0.015,\,0.025,\, 0.035$,
and $am_h=0.050,\, 0.060,\, 0.080,\, 0.100$. We used fun\-da\-men\-tal-adjoint gauge action 
with $\beta=4.0, \beta_a=-\beta/4$ \cite{Cheng:2013bca,Cheng:2013xha} and
nHYP-smeared staggered fermions \cite{Hasenfratz:2007rf}. Recently we 
also performed simulations with $\beta=4.4$. Lattice sizes were mostly 
$24^3 \times 48$ and $32^3 \times 64$, but also $16^3 \times 32$~(exploratory), 
$36^3 \times 64$ and $48^3 \times 96$.  The lattice generation was done
with the hybrid  Monte Carlo algorithm with one Hasenbusch intermediate 
mass; most simulations/measurements were performed with 
FUEL~\cite{Osborn:2014kda,FUEL}; most calculations were done with USQCD SciDAC software 
on USQCD computers at Fermilab and NSF-MRI computers at the MGHPCC.
Disconnected diagrams (for the $0^{++}$ singlet states) were
computed with stochastic sources (6 sources, full color and
time dilution, even-odd space dilution.)
\begin{figure}[t]
\centering
{\includegraphics[width=0.5\textwidth]{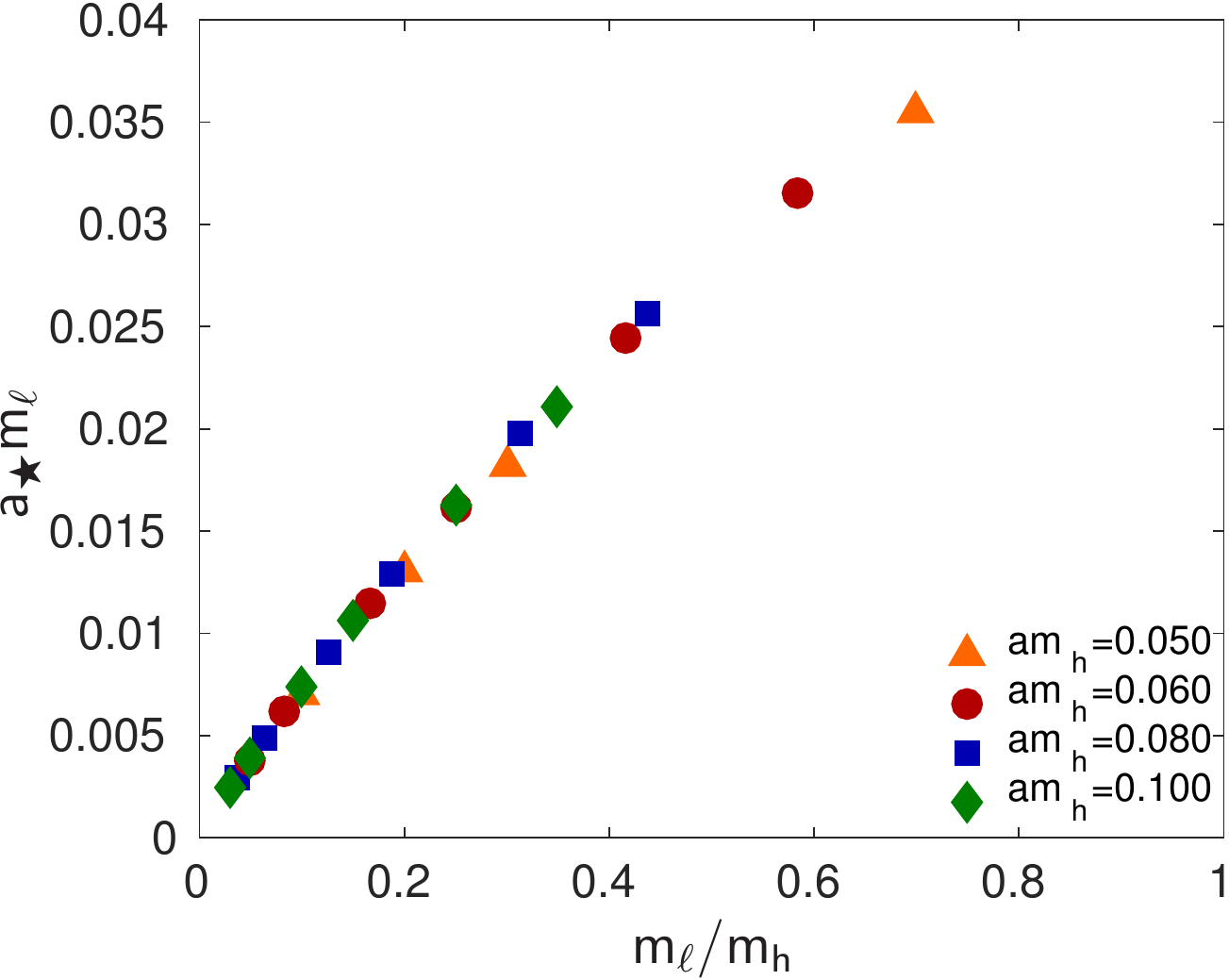}}
\caption{The masses $a_\star m_\ell$ displayed as function of the ratios $m_\ell/m_h$.
Universality implies that they should fall on a single curve.}
\label{fig:astar}
\end{figure}
We used the gradient flow method to set the scale for the various lattices.
The gradient flow defines the renormalized coupling $g_{\rm GF}$ by
$g_{GF}^2(\mu=1/\sqrt{8t}) = t^2\langle E(t)\rangle/{\cal N}$ where
$t$ is the flow time, $E(t)$ is the action~density and
${\cal N}$ is conventionally taken as $3(N_c-1)/(128 \pi^2)$
\cite{Narayanan:2006rf,Luscher:2009eq,Luscher:2010iy}.   We set the scale demanding
that $g_{GF}^2(t=t_0) = 0.3/{\cal N}$.
The lattice flow time $t_0/a^2$ then defines the scale. The symbol $a_\star$ appearing in some
of the following  figures  is used
 to provide a common scale for all dimension-full quantities.  It
denotes the lattice spacing for the simulation on 
the $36^3\times 64$ lattice with $am_\ell=0.003,\, am_h=0.080$.  The relation between $a_\star m_\ell$ and the 
ratios $m_\ell/m_h$ is worth noting.  As explained in the previous section, 
we expect dimensionless ratios of physical quantities to depend  on
the ratio $m_\ell/m_h$ but not individually on $m_\ell$ or $m_h$. Hyperscaling  manifests
itself also if quantities are plotted in terms of 
$a_\star m_\ell$ because  $a_\star$ can be considered as another physical quantity.\footnote{We recall that $m_\ell$ stands for the fixed physical
value of a light fermion mass.  Thus the values of $a_\star m_\ell$ 
are different from the quoted values for $a m_\ell$.  Unlike in QCD, the lattice spacing $a$ depends strongly on $m_\ell$ and $m_h$.}  The two variables $a_\star m_\ell$ and 
$m_\ell/m_h$ should therefore track each other falling on a single curve
and this is apparent in Fig~\ref{fig:astar}.

\begin{figure}[t]
  \centering
   \parbox{0.32\textwidth}{\includegraphics[height=0.28\textheight]{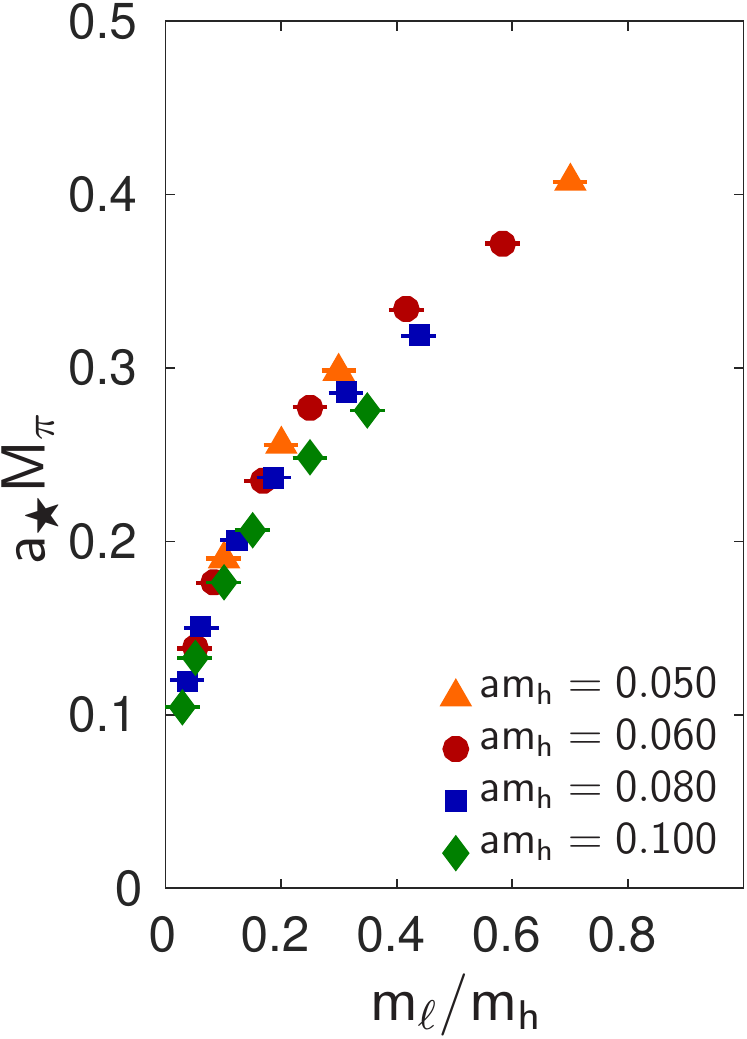}}  
   \parbox{0.32\textwidth}{\includegraphics[height=0.28\textheight]{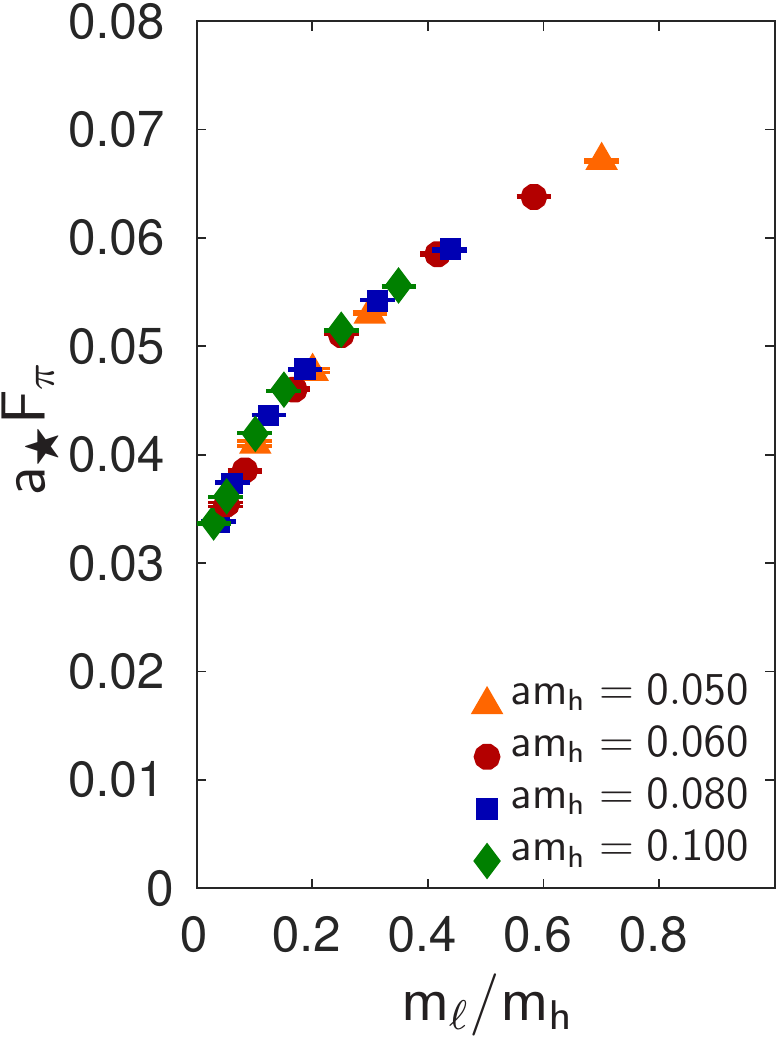}}
   \parbox{0.32\textwidth}{\includegraphics[height=0.28\textheight]{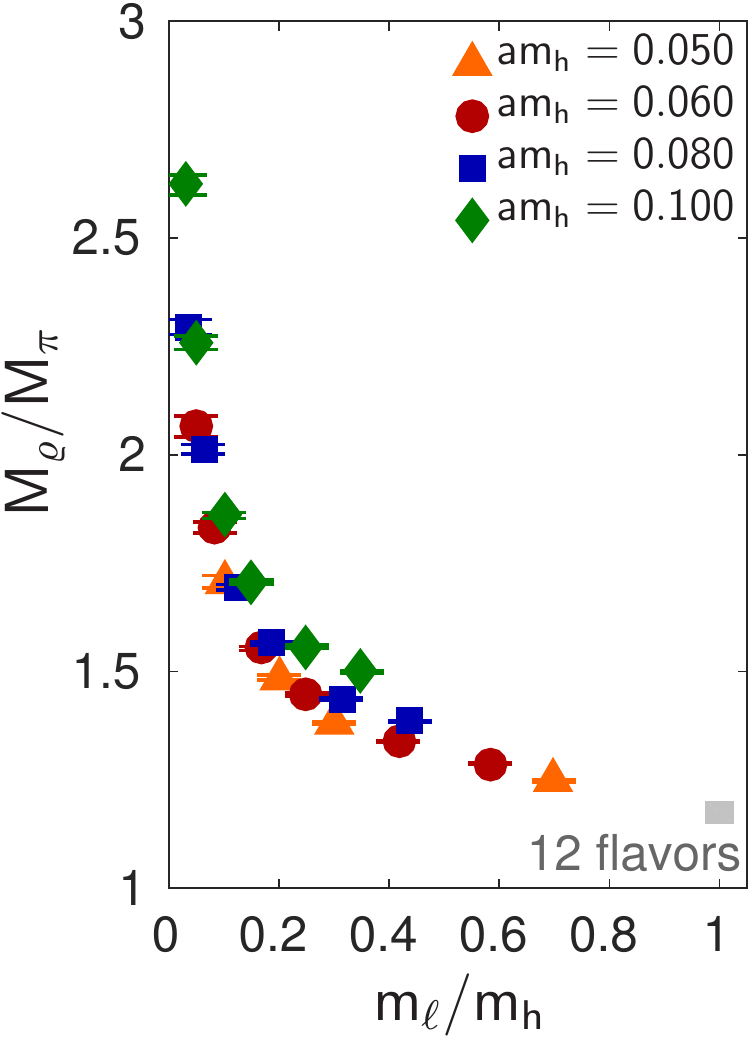}}
   
   \caption{The left panel shows the pseudoscalar meson mass $M_\pi$ and the panel in the middle its decay constant, $F_\pi$, as functions of $m_\ell/m_h$.  Note the universality as well 
as the fact that the pseudoscalar mass tends to zero in the chiral limit while $F_\pi$
tends to a finite limit, albeit with a steep drop.  The right panel shows
the ratio between the vector meson mass $m_\varrho$ and $m_\pi$, which diverges
in the chiral limit with chiral symmetry breaking, while it should tend to a
constant in a conformal theory. (We use hadron spectroscopy notation
to denote the states in our model and show only statistical errors.)
}
  \label{fig:ChirallyBroken}
\end{figure}
The first question one should address is whether the system with our parameter values exhibits chiral symmetry breaking and is not conformal.  Figure~\ref{fig:ChirallyBroken} provides 
strong evidence for that: in the chiral limit $M_\pi$ goes to zero, $F_\pi$ is finite, and $M_\varrho/M_\pi$ diverges.

\bigskip
\begin{figure}[ht]
  \centering
  \parbox{0.27\textwidth}{\includegraphics[height=0.24\textheight]{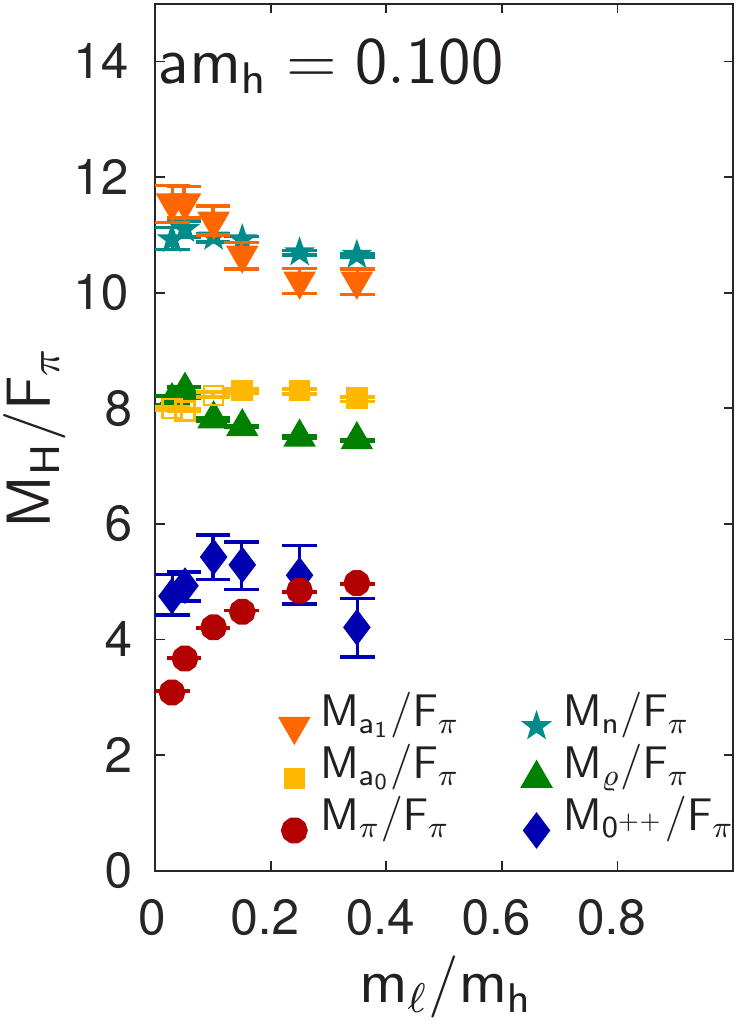}}
  \parbox{0.22\textwidth}{\includegraphics[height=0.24\textheight]{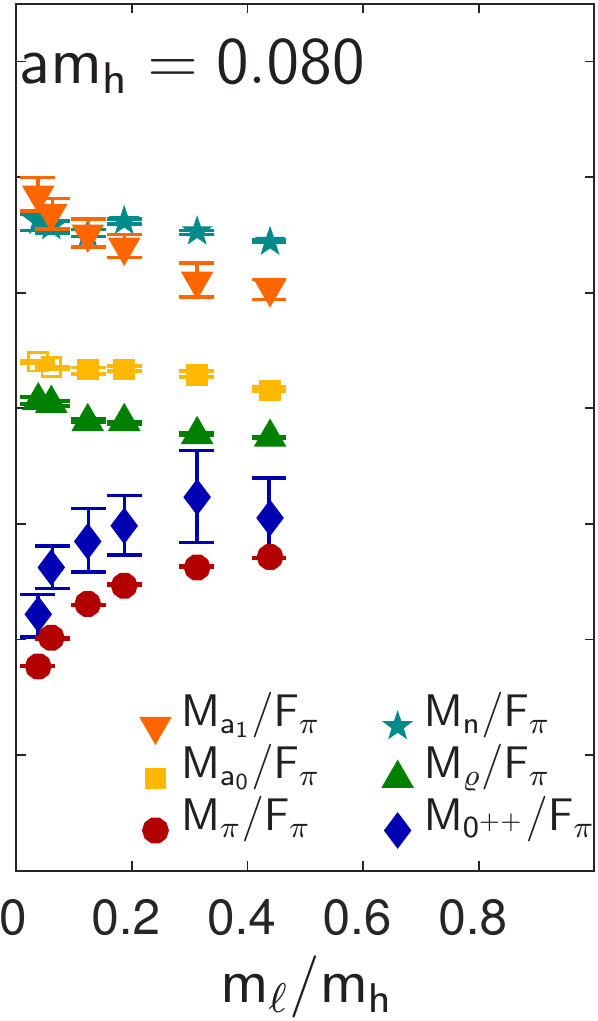}}
  \parbox{0.22\textwidth}{\includegraphics[height=0.24\textheight]{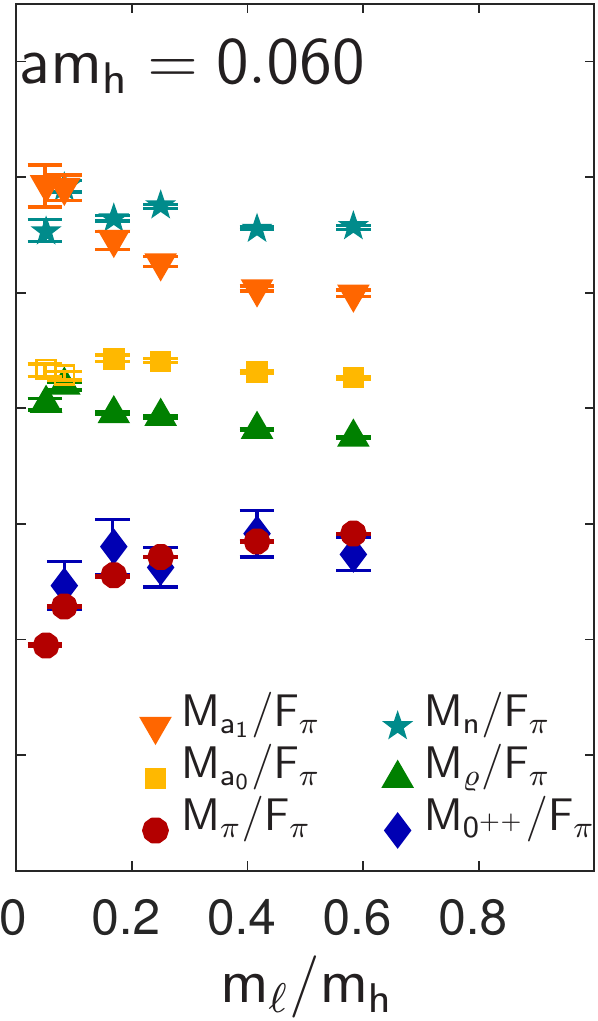}}
  \parbox{0.22\textwidth}{\includegraphics[height=0.24\textheight]{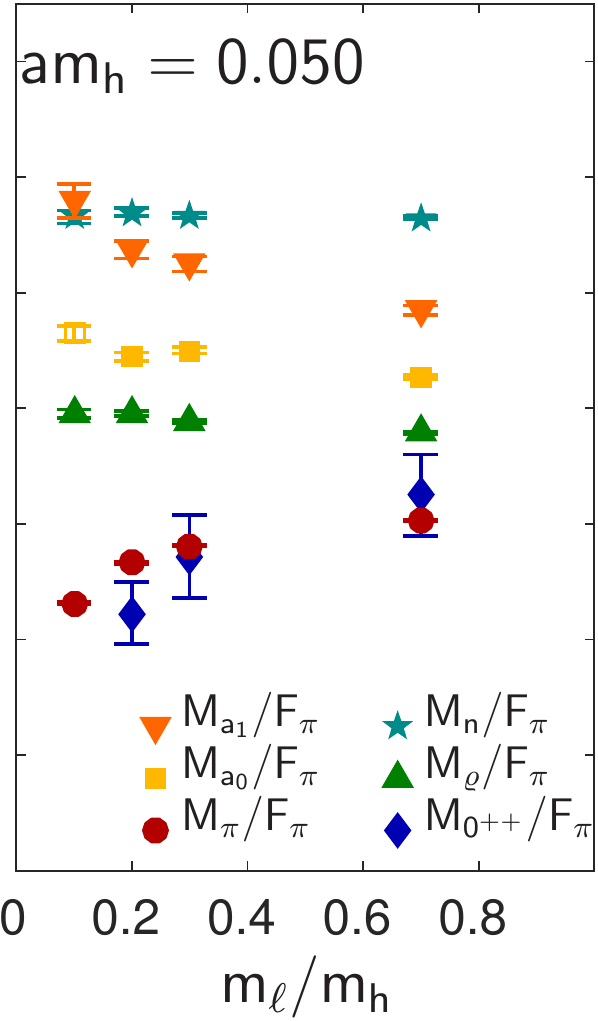}}
  \caption{The pion, rho, isosinglet $0^{++}$ and isomultiplet  $a_0$ scalar, axial,  and nucleon mass of the light-light flavor spectrum in units of $F_\pi$ (errors are statistical only). The four panels show the $N_f=4+8$ spectrum as the function of the ratio $m_\ell/m_h$ for $am_h = 0.100$, $0.080$, $0.060$, and $0.050$.  }
  \label{fig:Spectrum}
\end{figure}
Figure~\ref{fig:Spectrum} shows the masses of the light-light composites as dimensionless ratios over $F_\pi$ for
the four values of the heavy fermion masses, separately, plotted as a function 
of the dimensionless ratio $m_\ell/m_h$.  Superimposing the panels, which we do in a later
figure, would show hyperscaling, but here we keep the data apart for clarity 
of presentation. An interesting and possibly very important feature, already
noticed in~\cite{Brower:2015owo} and by other researchers, is that the mass of the
singlet scalar meson appears to track the pseudoscalar mass through the
range of light fermion masses.  If this were to continue down toward
the chiral limit, it would indicate that BSM theories based on strong dynamics
may give origin to a light scalar meson, as would be necessary for an
interpretation of the Higgs as a dilaton-like composite state. 

\begin{figure}[t]
  \centering
  {\parbox{0.083\textwidth}{\includegraphics[height=0.24\textheight]{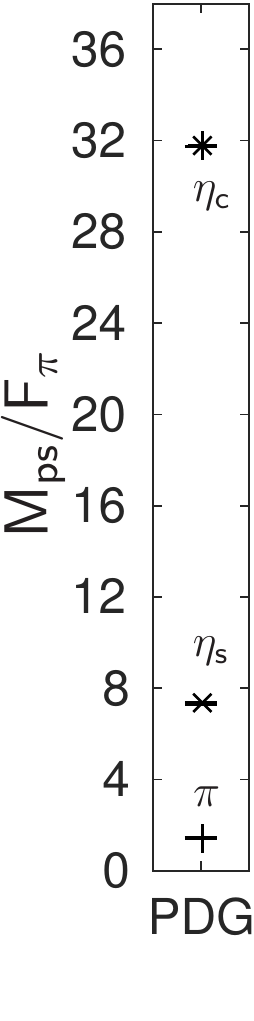}}  
  \parbox{0.198\textwidth}{\includegraphics[height=0.24\textheight]{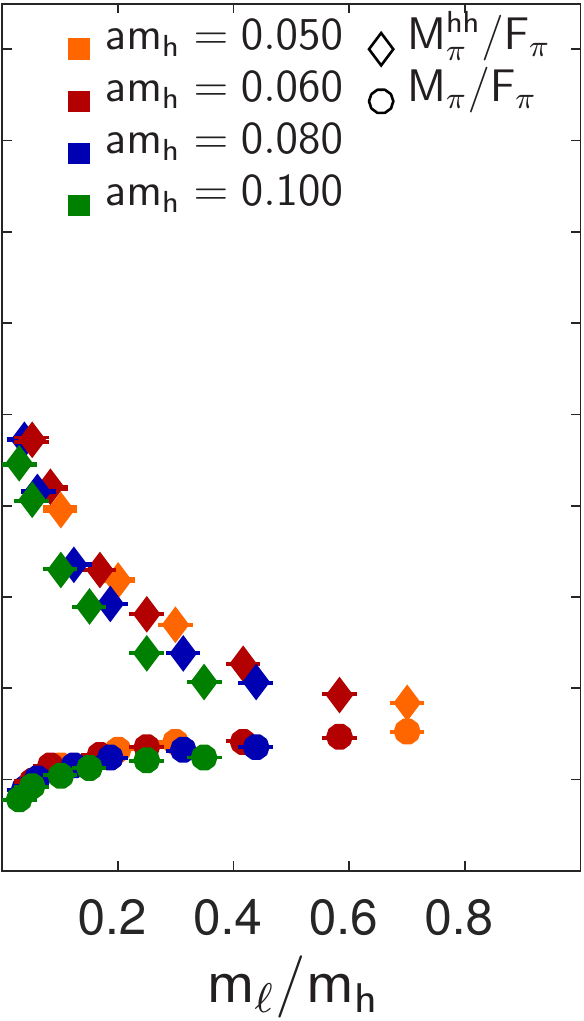}}
  \parbox{0.03\textwidth}{\includegraphics[height=0.24\textheight]{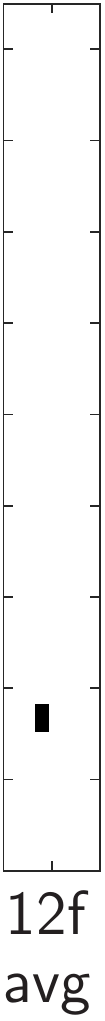}}}
\,
  \parbox{0.083\textwidth}{\includegraphics[height=0.24\textheight]{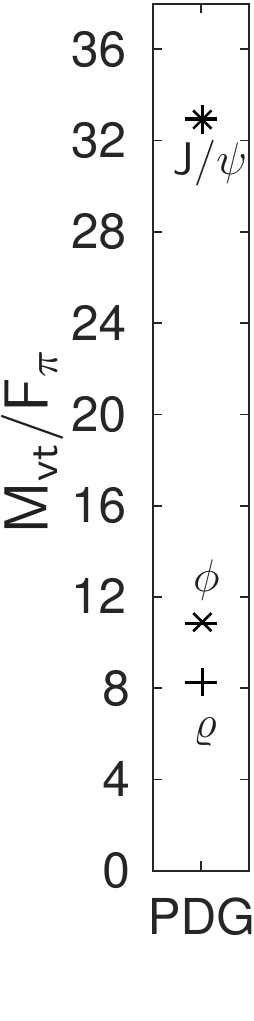}}  
  \parbox{0.198\textwidth}{\includegraphics[height=0.24\textheight]{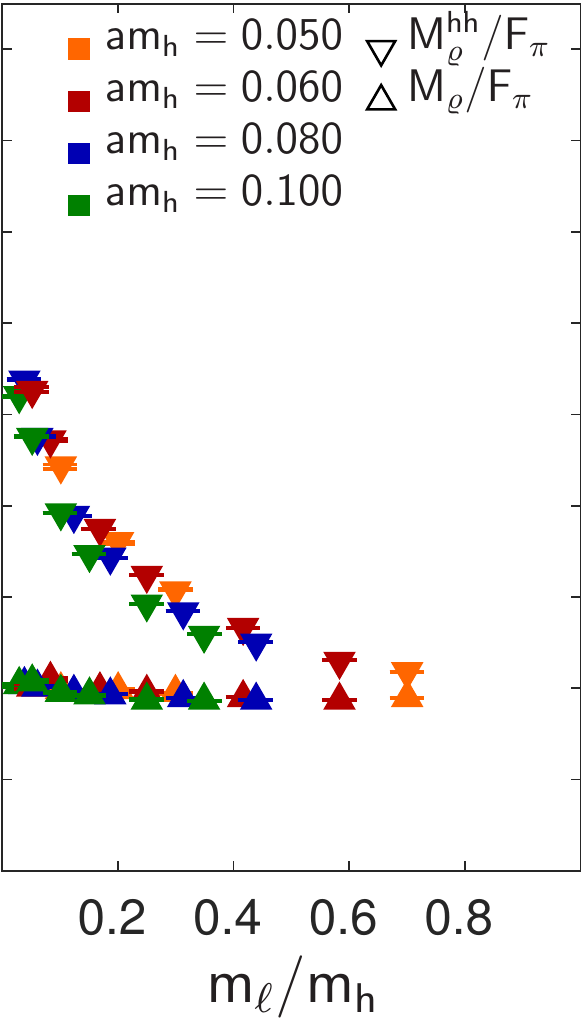}}
  \parbox{0.03\textwidth}{\includegraphics[height=0.24\textheight]{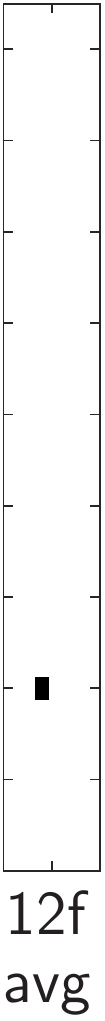}}
\,
  \parbox{0.083\textwidth}{\includegraphics[height=0.24\textheight]{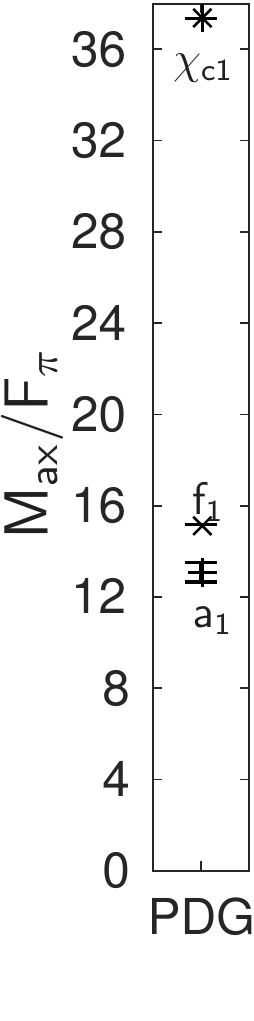}}  
  \parbox{0.198\textwidth}{\includegraphics[height=0.24\textheight]{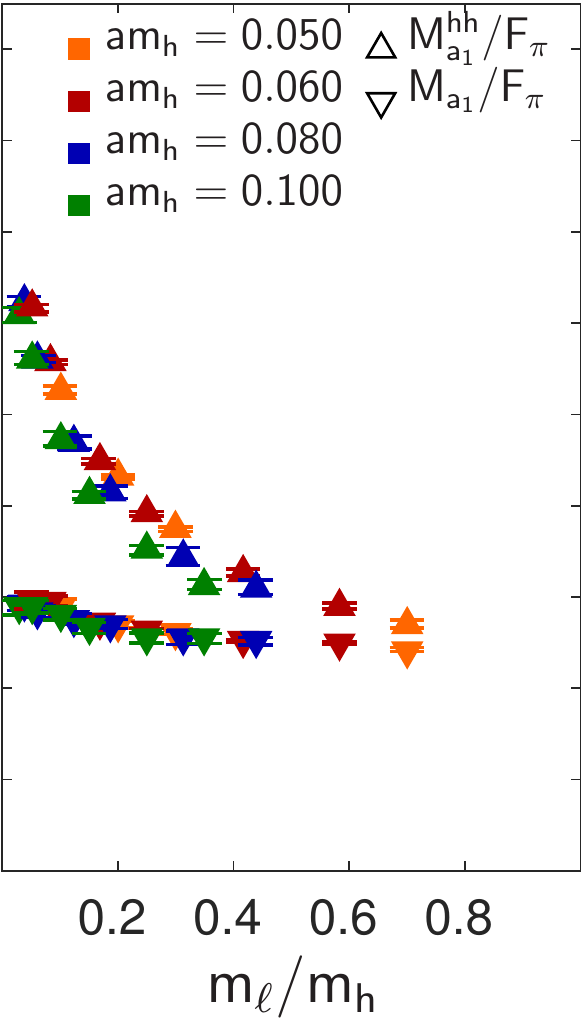}}
  \parbox{0.03\textwidth}{\includegraphics[height=0.24\textheight]{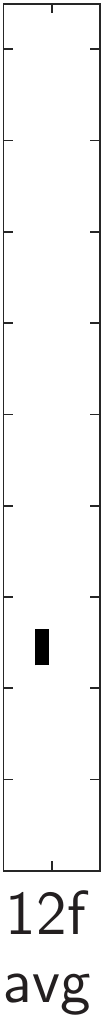}}  
  \caption{
    The three set of panels show  pseudoscalar (ps), vector (vt), and axial (ax) meson masses in units of $F_\pi$. The wide central panels show our data (with statistical errors only) as  function of $m_\ell/ m_h$ and different colors indicate the different $m_h$ values. The small panels to the right show averaged values for degenerate 12 flavors \cite{Aoki:2012eq,Fodor:2011tu,Cheng:2013xha,Aoki:2013zsa} and the panels on the left the corresponding PDG values \cite{Agashe:2014kda} for QCD divided by $F_\pi=94$ MeV. Values for the corresponding bottomonium states ($\eta_b$, $\Upsilon$, and $\chi_{b1}$) are too heavy to be shown on a reasonable scale. While the pseudoscalar and vector states are in general well understood in QCD, pure $(s\bar s)$ states do not occur in nature. For the $\eta_s$ mass, we use the lattice determination, $M_{\eta_s} = 688.5(2.2)$ MeV \cite{Dowdall:2013rya}, and quote for the vector and axial the PDG entries for the $\phi(1020)$ and $f_1(1420)$, respectively. Regardless of ambiguities in the QCD values, these plots highlight the different character of our heavy-heavy spectrum. Due to the presence of an IRFP, the system shows hyperscaling and we observe independence of the $m_h$, an unusual behavior in QCD standards.}
  \label{fig:ps_vt_ax}
\end{figure}

Figure~\ref{fig:ps_vt_ax} captures the essence of our new 
results.  The data for the masses of light-light and heavy-heavy pseudoscalar,
vector and axial vector states obtained with the four different heavy
fermion masses are all plotted together as function of $m_\ell/m_h$.
The data strikingly fall on universal curves as expected from hyperscaling.
The small deviations can be explained by remaining lattice artifacts and  scaling violations due to a slow approach to the IRFP.

\begin{figure}[t]
  \centering
   \parbox{0.4\textwidth}{\includegraphics[height=0.34\textheight]{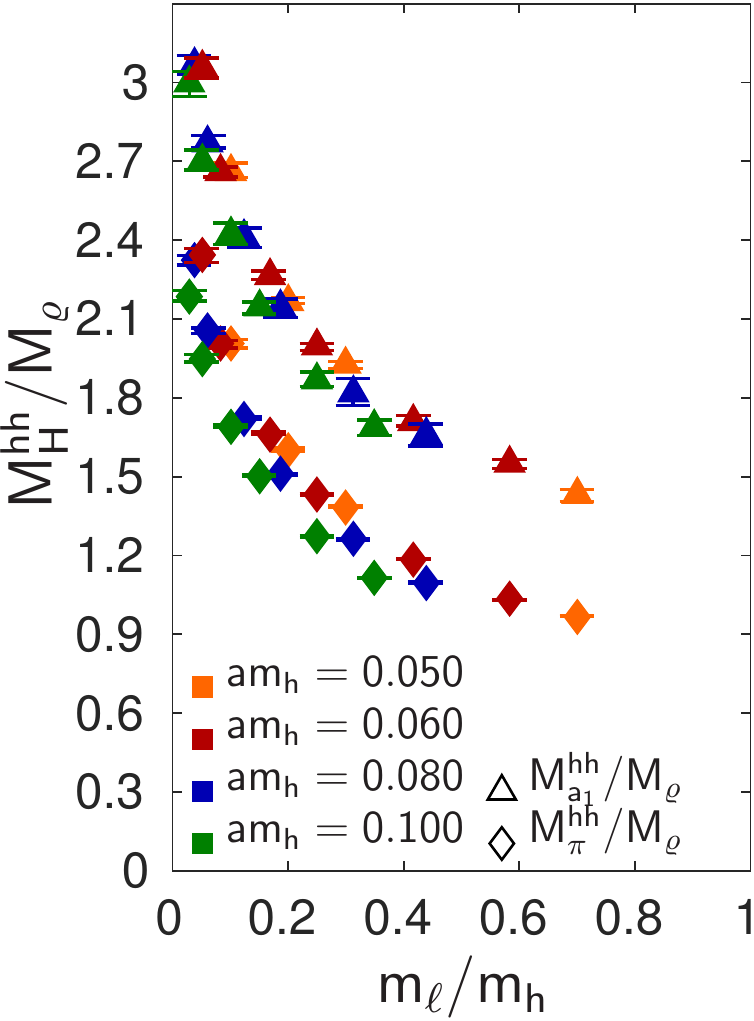}}
   \parbox{0.4\textwidth}{\includegraphics[height=0.34\textheight]{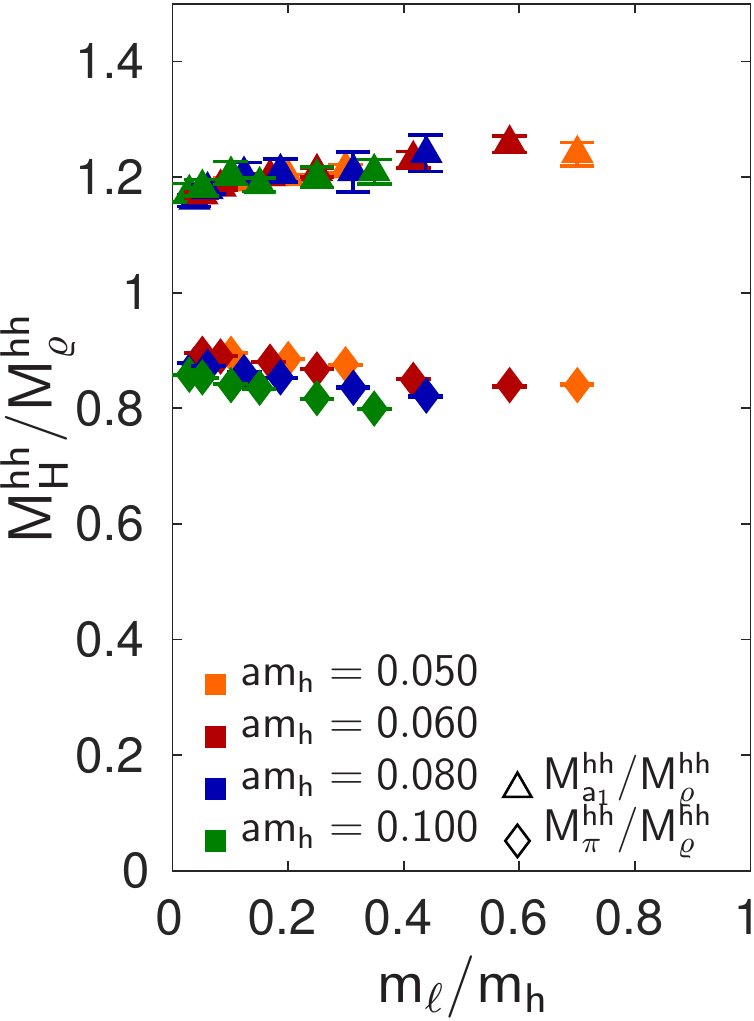}}
   \caption{The masses of the heavy-heavy $\pi$ and $a_1$ states (with statistical errors only)
       measured against the light-light (left panel) and heavy-heavy (right panel) $\varrho$ states.
}
  \label{fig:AlternativeRatios}
\end{figure}

The patterns of masses of the heavy-heavy composites shown in
 Fig.~\ref{fig:ps_vt_ax} bend noticeably upward as one approaches 
the light fermions'  chiral limit.  This is, however, to a large
extent a consequence of the fact that, in order to display dimensionless
quantities, we plot the ratios $M^{\rm hh}/F_\pi$.  As one can see from 
the  central panel of Fig.~\ref{fig:ChirallyBroken}, 
$F_\pi$ exhibits a rather sharp decrease
as one approaches the chiral limit and this causes an upward bend
of the ratios $M^{\rm hh}/F_\pi$.  In Figure~\ref{fig:AlternativeRatios} we
show the masses of the heavy-heavy $\pi$ and $a_1$ states 
measured against the light-light (left panel) and heavy-heavy (right panel)
$\varrho$ masses.  While the ratios  $M^{\rm hh}/M_\varrho$ still exhibit an upward
bend, due to the fact that, like $F_\pi$, the light-light $M_\varrho$ also bends 
downward as one approaches the chiral limit (this was not explicitly shown,
but can be inferred from the almost constancy of $M_\varrho/F_\pi$, see 
Fig.~\ref{fig:ps_vt_ax}), the ratios $M^{\rm hh}/M^{\rm hh}_\varrho$ show
little variation with $m_\ell/m_h$. The fact remains, though, that in 
phenomenological applications it would be the light-light $F_\pi$ which
sets the scale of the electroweak symmetry breaking and thus the increase
of the ratios $M^{\rm hh}/F_\pi$ would be of phenomenological relevance.
And this brings us to an important consideration, discussed in the following 
paragraph.

Our system is a prototype pseudo Nambu-Goldstone boson (PNGB) model (for recent work see e.g.~\cite{Contino:2010rs,Ferretti:2013kya,Vecchi:2015fma, Ma:2015gra}) in the
sense that, while in no way it can be a realistic model of EWSB, it exhibits
some of the main features of an eventual full fledged theory of EWSB based on 
strong dynamics: we expect that such a theory would have an IRFP in the limit
of massless fermions in order to allow for the required separation of scales, 
and that it would have extra fermions, beyond the light fermions giving origin 
to the NG bosons, to push it toward the conformal window.  Then one obvious
question is how the extra fermions should be handled and what would be the
implications of their existence.  Would their mass represent an additional
parameter in the theory?  Would they imply a spectrum of composites whose
masses will depend on the constituent fermion mass? and if so how should
this mass be tuned?  The answer that we gather from our investigation is
that, because of the IRFP, the spectrum of all composite states would
show hyperscaling.  As a consequence the spectrum of composites  would be
largely independent from the bare mass of the additional fermions.  Tuning
it  would only serve to change the range of walking.  The actual spectrum of 
composites  would of course depend on the underlying theory and it would 
constitute a set of excitations which  would show up at some point in
high energy collisions.

 Our final consideration is about our choice of the value 
$\beta=4.0$ for the gauge coupling constant.  We have argued that
$\beta$ is an irrelevant parameter and thus its specific value
should not matter.  Changing the gauge coupling near an IRFP is similar to changing the gauge action in a QCD simulation by adding irrelevant terms. The Wilson, Symanzik, Iwasaki or fixed point actions differ only in irrelevant terms and have the same continuum limit. In the same way conformal systems with different gauge coupling values near the IRFP  have the same continuum limit. Just like it is possible to choose a gauge action with reduced cutoff corrections in QCD, it is possible to choose  a $\beta$ value with reduced corrections to conformal hyperscaling. Both the finite size scaling analysis of Ref.~\cite{Cheng:2013xha}   and the step scaling function study of Ref.~\cite{Hasenfratz:2016dou} indicate that  $\beta\approx 4.5$ is optimal, showing minimal corrections to scaling in the $N_f=12$ system with our choice of gauge action.  At the same time larger $\beta$ values  correspond to a smaller lattice spacing  for the same bare mass, requiring simulations on larger volumes.   We chose $\beta = 4.0$ as a compromise after substantial
preliminary numerical exploration.

\begin{figure}[t]
  \centering
   \parbox{0.4\textwidth}{\includegraphics[height=0.34\textheight]{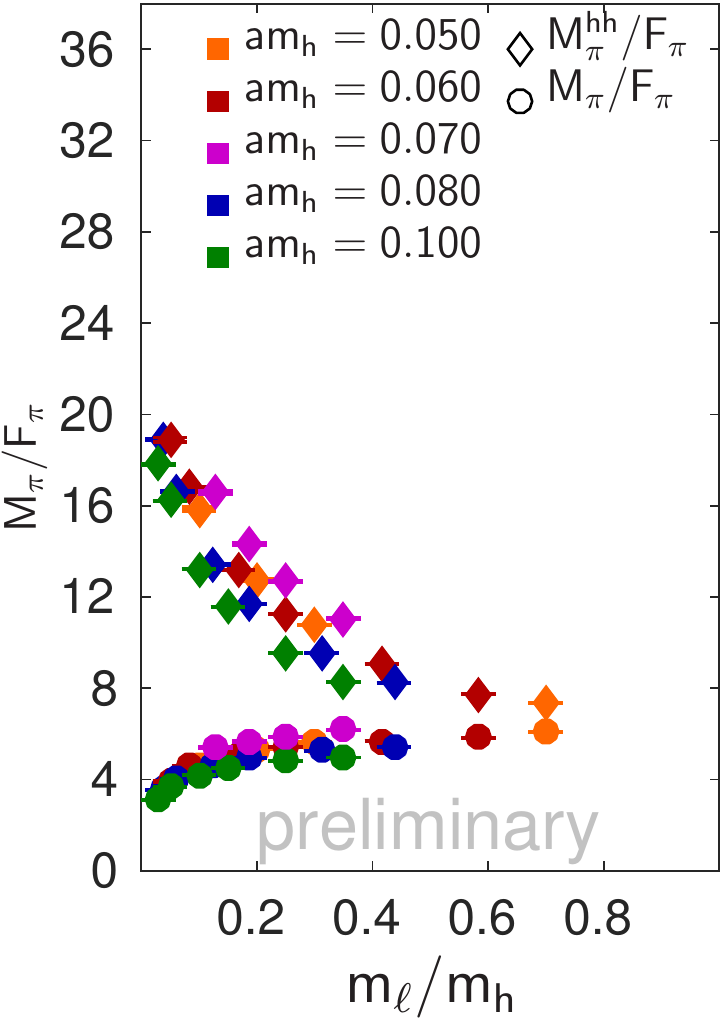}}
   \parbox{0.4\textwidth}{\includegraphics[height=0.34\textheight]{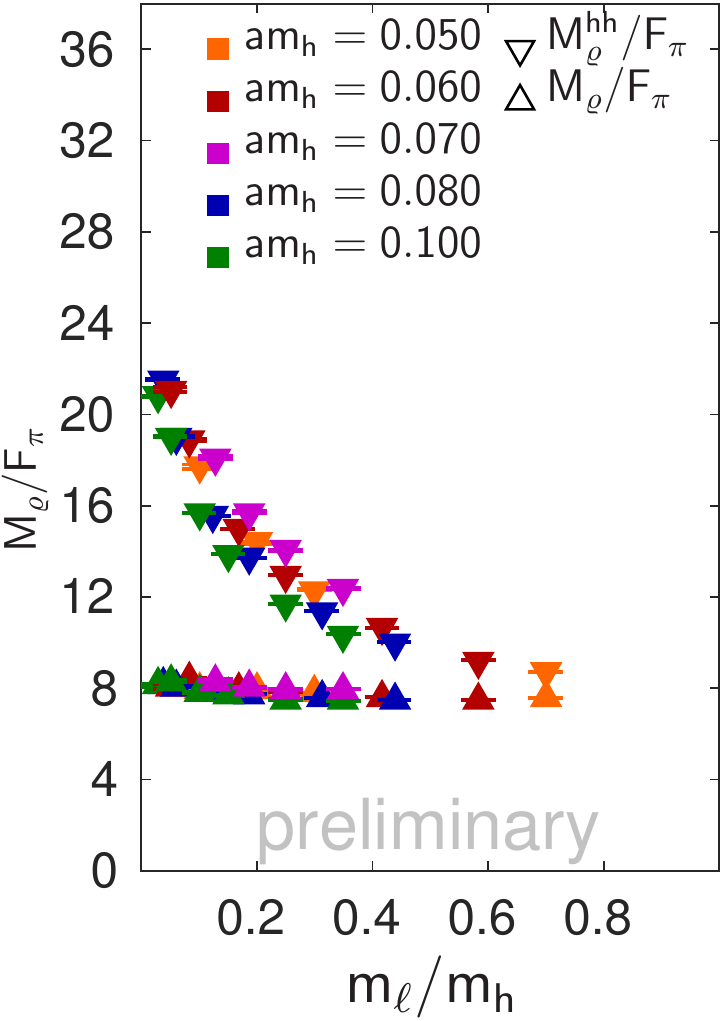}}
   \parbox{0.4\textwidth}{\includegraphics[height=0.34\textheight]{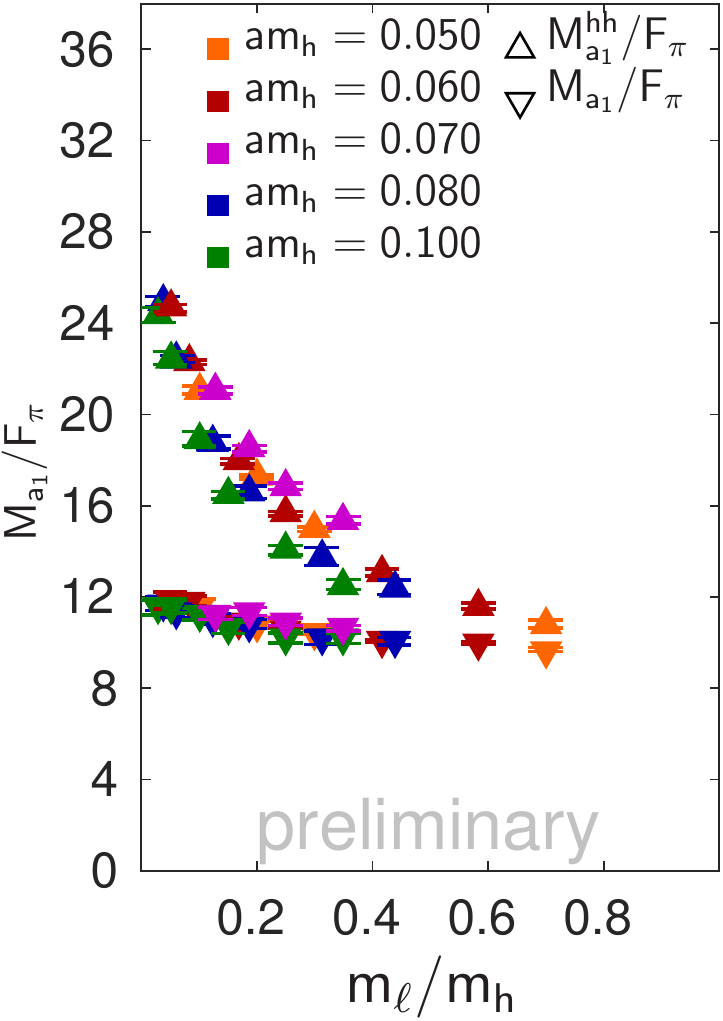}}   
   \caption{The same masses in units of $F_\pi$ as shown in Fig.~5
but supplemented here with preliminary data obtained with a different value of the bare
gauge coupling constant (the  purple points, with $\beta=4.4$ and $am_h=0.70$); all errors are statistical only.  
The fact that the
new data fall on the same curves as in Fig.~5 confirms that
the bare gauge coupling is an irrelevant parameter in the neighborhood of the
IRFP. }
  \label{fig:IrrelevantCoupling}
\end{figure}

Nonetheless we decided that it would be good to verify the expectation that $\beta$ is an irrelevant
parameter by running a simulation at a different value of the gauge 
coupling constant. Thus we recently  started simulations increasing $\beta$ by 10\%. Using  $32^3\times64$ volumes, $\beta=4.4$, and $am_h=0.070$ we are generating four ensembles with $am_\ell=0.009,\, 0.013125,\, 0.0175$, and $0.0245$. Carrying out spectrum measurements on the currently available gauge field configurations, we show our preliminary results for pseudoscalar, vector, and axial states by the  purple symbols in Fig.~\ref{fig:IrrelevantCoupling} and observe that the  purple symbols ($\beta=4.4$ data) follow closely the lines
traced by the $\beta=4.0$ data. This supports the notion that the
gauge coupling constant is an irrelevant parameter in the
neighborhood of the IRFP.  In the future, we plan to investigate finite volume effects in greater detail and will also study scaling violations more carefully by directly comparing data at different $\beta$-values but the same $m_\ell/m_h$ ratio.

\section{Conclusions}

We simulated an SU(3) gauge system with four light  and eight heavy
 flavors of variable mass to study the dependence of the spectrum
of composite excitations on the mass of the heavy  flavors.  We
found strong evidence of hyperscaling, with the masses of the light-light
and heavy-heavy composites falling on universal lines independently
of the heavy fermion mass.  We  emphasize this finding is solely based on analyzing dimenionless ratios and hence independent of how a lattice scale is defined. Although this was not the primary
purpose of our investigation, our results lend credence to the
notion that the SU(3) theory with 12 massless flavors has an
infrared fixed point.

Our results show that in our setup the heavy fermion mass plays 
a role similar to the role of the gauge coupling constant $\beta$
in QCD simulations.  In QCD $\beta$ is a (marginally) relevant
parameter and the QCD spectrum
is largely independent of $\beta$ (apart of course from discretization
errors). The only effect of a change of $\beta$  is a corresponding
change of scale.  Similarly, in our system a change of $m_h$ has no effect
on the spectrum of excitations (apart from higher order corrections)
and only modifies the length of the walking window; the gauge coupling $\beta$ is now an irrelevant parameter.  Since our model is based on a conformal fixed point where both the light and heavy fermion masses scale with the same anomalous dimension, physical quantities show hyperscaling and depend only on the ratio of fermion masses. This is  very restrictive and qualitatively different from  QCD.  Thus our model, and in general models based on a conformal fixed point, are an interesting alternative to describe chirally broken systems with unique properties.  

Finally, our investigation indicates that realistic strongly coupled 
BSM models,  if they  incorporate heavier 
mass fermions and  have a dynamics governed by an IRFP,   would exhibit
a spectrum of heavy-heavy excitations of great phenomenological 
interest. Even more interesting are  heavy-light states because  they would couple directly to  SM particles through the light fermions. While we have not investigated the heavy-light spectrum, those states are  expected to lie between the light-light and heavy-heavy spectrum, making them accessible to experiments if BSM physics is described by a model similar to ours.
\newline
\section*{Acknowledgments}
\label{acknowledgements}
The authors thank their colleagues in the LSD Collaboration for fruitful and inspiring discussions.
Computations for this work were carried out in part on facilities of the USQCD Collaboration, which are funded by the Office of Science of the U.S.~Department of Energy, on computers at the MGHPCC, in part funded by the National Science Foundation, and on computers allocated under the NSF Xsede program to the project TG-PHY120002.
We thank Boston University, Fermilab, the NSF and the U.S.~DOE for providing the facilities essential for the completion of this work.  A.H. acknowledges support by DOE grant DE-SC0010005 and C.R. by DOE grant DE-SC0015845.   This project has received funding from the European Union's Horizon 2020 research and innovation programme under the Marie Sk{\l}odowska-Curie grant agreement No 659322.

{\small
  \bibliography{../General/BSM}
  \bibliographystyle{apsrev4-1}
}


\end{document}